\RequirePackage{lineno}
\documentclass[twocolumn,aps,prl,showpacs,superscriptaddress,floatfix]{revtex4-1}
\usepackage{epsfig}
\usepackage{hyperref}
\usepackage{lineno}
\usepackage{url}
\usepackage{multirow}
\usepackage{xcolor}
\usepackage{float}
\usepackage{rotating}

\usepackage{amsmath}
\usepackage{graphicx}
\usepackage[caption=false]{subfig}

\usepackage{multirow}
\usepackage{adjustbox}
\usepackage{cellspace}
\usepackage{makecell}

\makeatletter
\renewcommand{\p@subsection}{}
\renewcommand{\p@subsubsection}{}
\makeatother

\makeatletter
\let\LN@equation\equation
\let\LN@endequation\endequation
\renewcommand{\equation}{\linenomath\LN@equation}
\renewcommand{\endequation}{\LN@endequation\endlinenomath}
\let\LN@gather\gather
\let\LN@endgather\endgather
\renewcommand{\gather}{\linenomath\LN@gather}
\renewcommand{\endgather}{\LN@endgather\endlinenomath}
\makeatother

\begin{document}

\newcommand{\rp}{\Psi_{\text{SP}}}
\newcommand{\pp}{\Psi_{\text{PP}}}
\newcommand{\fcme}{f_{\text{CME}}}

\title{Comment on ``A sensitivity study of the primary correlators used to characterize chiral-magnetically-driven charge separation'' by Magdy, Nie, Ma, and Lacey}

\author{Yicheng Feng}
\address{Department of Physics and Astronomy, Purdue University, West Lafayette, IN 47907, USA}

\author{Fuqiang Wang}
\email{All correspondence should be addressed to F.W. $<$fqwang@purdue.edu$>$.}
\address{Department of Physics and Astronomy, Purdue University, West Lafayette, IN 47907, USA}

\author{Jie Zhao}
\address{Department of Physics and Astronomy, Purdue University, West Lafayette, IN 47907, USA}

\date{\today} 


\begin{abstract}
  This note points out an apparent error in the publication Phys.~Lett.~B {\bf 809} (2020) 135771 by Magdy, Nie, Ma, and Lacey.
\end{abstract}

\maketitle


This note concerns an apparent error in the statistical uncertainties in 
``A sensitivity study of the primary correlators used to characterize chiral-magnetically-driven charge separation''
by Magdy, Nie, Ma, and Lacey,
published in Phys.~Lett.~B {\bf 809} (2020) 135771 (MNML).

Table~\ref{TabX} lists the data points read off from 
Fig.~2 ($\Delta\gamma(\rp)$, $\Delta\gamma(\pp)$)
and Fig.~3 ($f_1$, $f_2$, $\fcme$) of MNML 
by a digital ruler (\url{https://apps.automeris.io/wpd/}).
The quantities $\Delta\gamma(\rp)$ and $\Delta\gamma(\pp)$ are the charge-dependent azimuthal correlators~\cite{Voloshin} with respect to the spectator plane (SP) and the participant plane (PP), respectively, in 10-50\% centrality Au+Au collisions simulated by the AMPT (A Multi-Phase Transport) model in MNML.
Using conventions in MNML,
$r_1 = \Delta\gamma(\rp) / \Delta\gamma(\pp)$,
$r_2 = v_2(\rp) / v_2(\pp)$ (where $v_2(\rp)$ and $v_2(\pp)$ are the elliptic flow parameters with respect to SP and PP, respectively),
$f_1 = r_1 / r_2 -1$, and
$f_2 = 1 / r_2^2 -1$,
the chiral magnetic effect (CME) signal fraction in the $\Delta\gamma(\pp)$ measurement is given by $\fcme = f_1 / f_2$~\cite{Xu:2018cpc}. The SP and RP (reaction plane) were interchangeable in those formulas, and SP was used in the calculations as stated in MNML.
From the read-off data points, we compute $r_1$, assuming uncorrelated $\Delta\gamma(\rp)$ and $\Delta\gamma(\pp)$ errors; since $r_2$ is not readily accessible from MNML, we compute it by $r_{2} = 1/\sqrt{f_{2}+1}$; from $r_1$ and $r_2$ we compute $f_1$, and then $\fcme$ using our computed $f_1$.
These quantities are also listed in Table~\ref{TabX}.

Our computed errors on $f_1$ and $\fcme$ are {\em many} times larger than those in MNML, depending on the values of $f_1$. The absolute error on $f_1$ is of course equal to the absolute error on $r_1/r_2$.
Since $f_1$ can be zero, the relative error on $f_1$ can blow up (cf. Table~\ref{TabX}).
It appears, however, that the {\em relative} error on $f_1$ in MNML equals approximately to the {\em relative} error on $r_1$ in Table~\ref{TabX}. (Note that the digital ruler could introduce some imprecision in the read-out numbers. Also note that the error on $r_2$ is negligible compared to that on $r_1$; 
whether or not the $r_2$ error was properly propagated to $f_2$ in MNML, 
which in turn affects our calculated $r_2$, is of no significance.)
If the relative error on $r_1/r_2$ was mistaken as the $f_1$ relative error in MNML, then the $f_1$ absolute error could be very small when $f_1\sim0$; our digital ruler failed to read the errors of the two $f_1\sim0$ data points in MNML.

This issue of the apparent incorrect errors was pointed out by us to two of the authors of MNML (Magdy, Lacey) 
at an internal physics discussion meeting in STAR (\url{https://www.star.bnl.gov/}) when the preprint version (arXiv:2002.07934v1) of MNML appeared. 
It was also pointed out at the meeting, by examining the relevant analysis code, that there was a double counting of particle pairs, artificially reducing the statistical errors by a factor of $\sqrt{2}$; this was acknowledged by the authors at the STAR discussion meeting. Since this cannot be verified with the information available in MNML, we do not consider it here; considering it would increase all the errors by factor $\sqrt{2}$.
When a newer preprint (Lacey and Magdy, arXiv:2006.04132v2) later appeared, which had the same $\fcme$ data points, we pointed out the issue to the authors again, also at a STAR meeting. Despite of the multiple remonstrations, the issue was not fixed; the data points published in MNML are identical to those in the arXiv preprints.

\begin{figure}
  \includegraphics[width=1.00\linewidth]{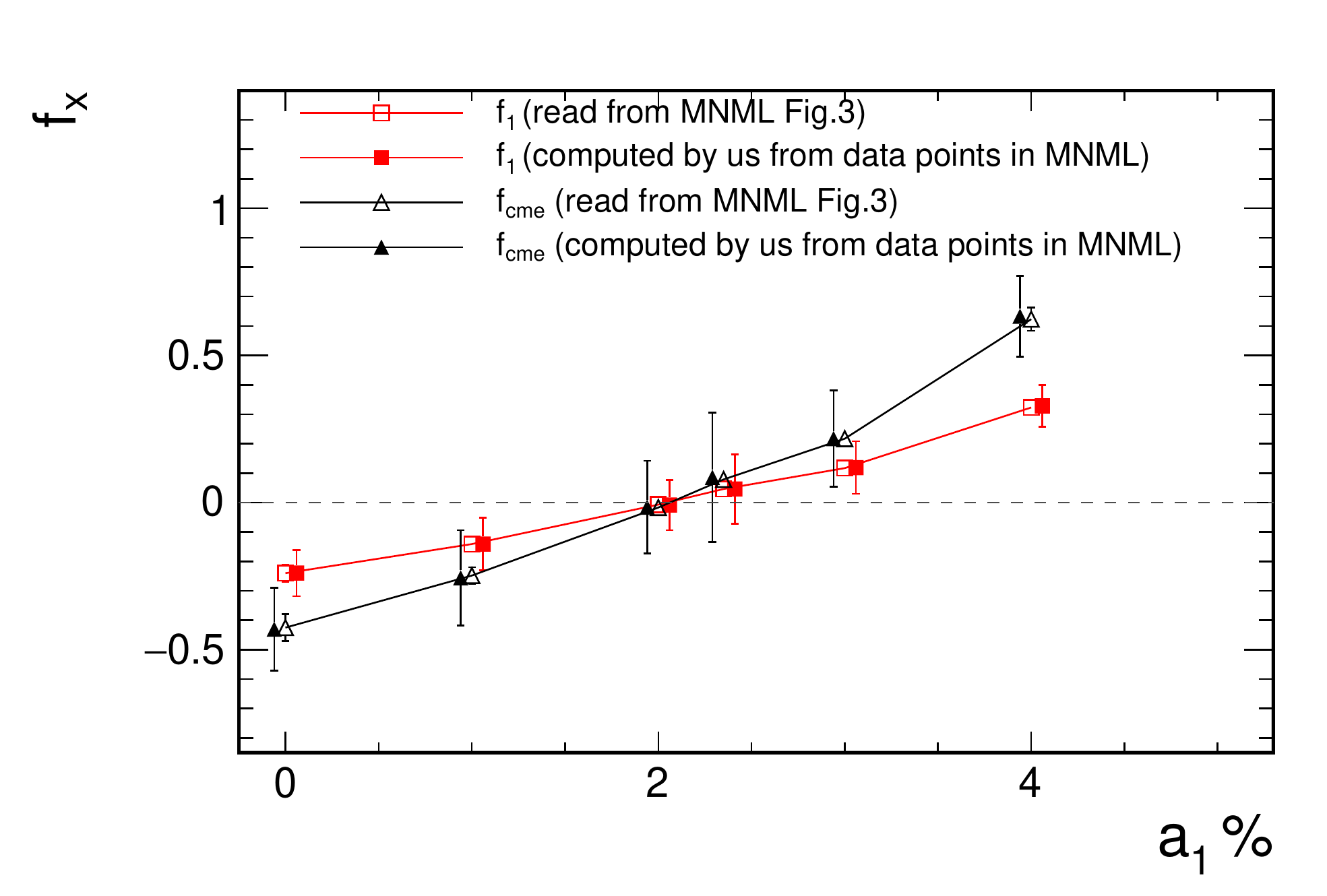}
  \caption{The $f_1$ and $\fcme$ as functions of $a_1$, the CME signal input to AMPT in MNML. The hollow markers are those read from MNML Fig.~3. The solid markers are those computed by us with proper error propagation.}
  \label{FigFx}
\end{figure}

Figure~\ref{FigFx} depicts our computed $f_1$ and $\fcme$ in solid markers, compared to those from MNML in hollow markers. 
With our correctly propagated errors, the data points (solid markers) appear to be too smooth, relative to the error bars.
Fitting a quadratic function to our computed $f_1$ gives a $\chi^2/\text{NDF}=0.074/3$ and a p-value of $0.995$ (i.e.~the probability for a lower $\chi^2/\text{NDF}$ value is 0.5\%; if the errors were already artifically reduced by factor $\sqrt{2}$ because of a double counting in MNML, then the likelihood would be even smaller).
Fitting $\fcme$ gives similar result, as expected, because $\fcme$ is $f_1$ scaled by the essentially error-free $f_2$.
(Incidentally, a quadratic fit to the $f_1$ from MNML, with the {\em incorrect} errors, gives a numerically reasonable $\chi^2/\text{NDF}$ and p-value.) 
The error we computed for $f_1$ is predominately determined by the error on $r_1$. As expected, a quadratic fit to $r_1$ gives a $\chi^2/\text{NDF}=0.109/3$ and p-value of 0.991, similar to those for our computed $f_1$.
However, quadratic fits to the individual $\Delta\gamma(\rp)$ and $\Delta\gamma(\pp)$ give reasonable $\chi^2/\text{NDF}$ (p-value) of 5.22/3 (0.157) and 0.90/3 (0.825), respectively.
Since $r_1$ is the ratio of $\Delta\gamma(\rp)$ over $\Delta\gamma(\pp)$, one is forced to conclude that either the two $\Delta\gamma$ quantities are strongly correlated so standard error propagation does not apply or something is unnatural with the AMPT $\Delta\gamma$ data points in MNML.
For the former, in order for the $r_1$ error to be inflated by a factor of $\sim5$ (so that the fit $\chi^2/\text{NDF}\sim1$) from simple error propagation, the $\Delta\gamma(\rp)$ and $\Delta\gamma(\pp)$ need to be $\sqrt{1-(1/5)^2)}=98\%$ correlated if they have the same relative errors; since the error on $\Delta\gamma(\pp)$ is significantly larger than that on $\Delta\gamma(\rp)$, even if they were 100\% correlated, the $r_1$ error would not be factor 5 smaller than that from simple error propagation. Therefore, we conclude that the AMPT $\Delta\gamma$ data points in MNML are unnatural.

A few remarks are in order:
\begin{itemize}
\item
The authors of MNML make the point of a turn-on threshold effect in $\fcme$, obtained from the method utilizing the SP and PP first proposed in Ref.~\cite{Xu:2018cpc}.
With the correctly propagated errors, this point becomes moot.
\item
Fig.4(a) of MNML shows a convex $R_{\Psi_{2}}$ distribution from AMPT with no input CME signal ($a_1=0$).
This is contrary to other background studies using hydrodynamics~\cite{Bozek:2018aad}, toy model resonance simulations~\cite{Feng:2018so}, and AMPT of multiple versions~\cite{AMPT}.
\item
A non-flat $R_{\Psi_{2}}$ distribution, either convex or concave, 
means that $R_{\Psi_{2}}$ {\em is} sensitive to background. 
The convexity of the AMPT result with $a_{1}=0$ is comparable to the concavity of the $a_{1}=2\%$ result in MNML.
Omitting the $a_{1}=0$ point from MNML Fig.4(f), 
extrapolating only the $a_{1}>0$ points to a seeming zero intercept, 
hence claiming little background contamination in $R_{\Psi_{2}}$, is improper.
\end{itemize}

In conclusion, there is an apparent error in the statistical uncertainties in 
``A sensitivity study of the primary correlators used to characterize chiral-magnetically-driven charge separation''
by Magdy, Nie, Ma, and Lacey,
published in Phys.~Lett.~B {\bf 809} (2020) 135771.
This was pointed out by us to Magdy and Lacey, two of the authors, at an internal STAR meeting when the preprint version (arXiv:2002.07934v1) of the said publication appeared, and also later when another preprint (arXiv:2006.04132v2) using the same data points was posted. The apparent error was not fixed--the relevant data points published in MNML are identical to those in the arXiv preprints. Tracing this error reveals that the AMPT data points in MNML are statistically unnatural.

\begin{turnpage}
\begin{table*}[]
\caption{The $a_1$ values plotted in MNML Figs.~2 and 3 are slightly offset compared to the texts written in MNML Fig.~4, except one plotted at $a_1=2.35$\%. The $\Delta\gamma(\rp)$, $\Delta\gamma(\pp)$, $f_{1}$, $f_{2}$, and $\fcme$ values (middle black) are read from MNML Figs.~2 and 3 by a digital ruler (\url{https://apps.automeris.io/wpd/}). The $r_1$, $r_2$, $f_1$, and $\fcme$ (lower block) are computed by us with proper error propagation. The numbers in parentheses are relative errors for easy comparison.}
\begin{tabular}{ccccccc}
\hline
$a_{1}$ & 0\% & 1\% & 2\% & 2.35\% & 3\% & 4\% \\\hline
\multicolumn{6}{c}{Read from Figs.~2 and 3 of MNML}\\
$\Delta\gamma(\rp)\times10^5$ & $5.43\pm0.33(6.1\%)$ & $5.86\pm0.32(5.5\%)$ & $7.91\pm0.32(4.0\%)$ & $9.30\pm0.47(5.1\%)$ & $10.05\pm0.37(3.7\%)$ & $14.80\pm0.27(1.8\%)$ \\
$\Delta\gamma(\pp)\times10^5$ & $8.91\pm0.74(8.3\%)$ & $8.50\pm0.75(8.8\%)$ & $9.90\pm0.75(7.6\%)$ & $11.0\pm1.1(10\%)$ & $11.17\pm0.79(7.1\%)$ & $13.73\pm0.69(5.0\%)$ \\
$f_1$ (MNML) & $-0.240\pm0.029(12\%)$ & $-0.141\pm0.014(9.9\%)$ & $-0.007$ (error unreadable) & $0.046$ (error unreadable) & $0.117\pm0.011(9.4\%)$ & $0.323\pm0.021(6.5\%)$ \\
$f_2$ & $0.557\pm0.007(1.3\%)$ & $0.552\pm0.007(1.3\%)$ & $0.540\pm0.007(1.3\%)$ & $0.536\pm0.007(1.3\%)$ & $0.546\pm0.007(1.3\%)$ & $0.519\pm0.004(0.8\%)$ \\
$\fcme$ (MNML) & $-0.425\pm0.046(11\%)$ & $-0.248\pm0.028(11\%)$ & $-0.017$ (error unreadable) & $0.079\pm0.011(14\%)$ & $0.217\pm0.018(8.3\%)$ & $0.623\pm0.039(6.3\%)$ \\\hline
\multicolumn{6}{c}{Computed by us}\\
$r_1=\Delta\gamma(\rp)/\Delta\gamma(\pp)$ & $0.61\pm0.06(9.8\%)$ & $0.69\pm0.07(10\%)$ & $0.80\pm0.07(8.8\%)$ & $0.84\pm0.10(12\%)$ & $0.90\pm0.07(7.9\%)$ & $1.08\pm0.06(5.6\%)$ \\
$r_2=1/\sqrt{f_2+1}$ & $0.801\pm0.002(0.25\%)$ & $0.803\pm0.002(0.25\%)$ & $0.806\pm0.002(0.25\%)$ & $0.807\pm0.002(0.25\%)$ & $0.804\pm0.002(0.25\%)$ & $0.811\pm0.001(0.12\%)$ \\
$f_1=r_1/r_2-1$ & $-0.24\pm0.08(33\%)$ & $-0.14\pm0.09(64\%)$ & $-0.01\pm0.09(900\%)$ & $0.05\pm0.12(240\%)$ & $0.12\pm0.09(75\%)$ & $0.33\pm0.07(21\%)$ \\
$\fcme=f_1/f_2$ & $-0.43\pm0.14(33\%)$ & $-0.26\pm0.16(64\%)$ & $-0.02\pm0.16(900\%)$ & $0.09\pm0.22(240\%)$ & $0.22\pm0.16(75\%)$ & $0.63\pm0.14(21\%)$ \\\hline
\end{tabular}
\label{TabX}
\end{table*}
\end{turnpage}

This work was supported by the U.S.~Department of Energy under Grant No.~DE-SC0012910.

\end{document}